\documentclass[usegraphicx,fleqn]{mn2e}
\usepackage{times}
\usepackage{graphicx}
\usepackage{soul}	
\def\Msolar{\hbox{${\rm M}_\odot$}}

\title{Pre-main sequence stars older than 8\,Myr in the Eagle Nebula}

\author[G. De Marchi,  N. Panagia, M. G. Guarcello, R. Bonito]
{Guido~De~Marchi,$^1$ Nino~Panagia,$^{2,3,4}$ M.~G.~Guarcello,$^5$ 
and Rosaria~Bonito$^6$\\
$^1$European Space Research and Technology Centre, Keplerlaan 1, 2200 AG 
Noordwijk, Netherlands, gdemarchi@rssd.esa.int \\
$^2$Space Telescope Science Institute, 3700 San Martin Drive, Baltimore, MD
21218, USA, panagia@stsci.edu\\
$^3$INAF--NA, Osservatorio Astronomico di Capodimonte, Salita Moiariello
16, 80131 Naples, Italy\\
$^4$Supernova Ltd, OYV \#131, Northsound Rd., Virgin Gorda,  
British Virgin Islands, VG 1155\\
$^5$Smithsonian Astrophysical Observatory, 60 Garden Street, Cambridge,
MA 02138, USA, mguarcel@head.cfa.harvard.edu\\
$^6$Dipartimento di Fisica e Chimica, Universit\`a di Palermo and 
INAF--PA, Osservatorio Astronomico di Palermo, Piazza del \\ 
\hspace*{0.15cm}Parlamento 1, 90134 Palermo, Italy, sbonito@astropa.unipa.it
}

\begin{document}

\date{Received 20 June 2013; Accepted 30 July 2013}

\pagerange{\pageref{firstpage}--\pageref{lastpage}} \pubyear{2013}

\maketitle

\begin{abstract}

Attention is given to a population of 110 stars in the NGC\,6611 cluster
of the Eagle Nebula  that have prominent near-infrared (NIR) excess and
optical colours typical of pre-main sequence (PMS) stars older than
8\,Myr. At least half of those for which spectroscopy
exists have a H$\alpha$ emission line profile revealing active
accretion. In principle, the $V-I$  colours of all these stars would be
consistent with those of young PMS  objects ($< 1$\,Myr) whose radiation
is heavily obscured  by a circumstellar disc seen at high inclination
and in small part scattered towards the observer by the back side of the
disc. However, using theoretical models it is shown here that objects of
this type can only account for a few percent of this population. In
fact, the spatial distribution of these objects, their X-ray
luminosities, their optical brightness, their positions in the
colour--magnitude diagram  and the weak Li absorption lines of the stars
studied spectroscopically suggest that most of them are  at least 8
times older than the $\sim 1$\,Myr-old PMS stars already known in this
cluster and could be as old as $\sim 30$\,Myr. This is the largest
homogeneous sample to date of Galactic PMS stars considerably older than
8\,Myr that are still actively accreting from a circumstellar disc and
it allows us to set a lower limit of 7\,\% to the disc frequency at
$\sim 16$\,Myr in NGC\,6611. These values imply a characteristic
exponential lifetime of $\sim 6$\,Myr for disc dissipation.

\end{abstract}


\section{Introduction}

In the current paradigm of star formation (see e.g. Lada 1999 and
references therein) conservation of angular momentum during the collapse
of cloud cores leads to the formation of circumstellar discs around
newly born stars. The presence and evolution of circumstellar discs is
important both for planets, which find in discs the natural sites for
their formation, and for the star itself, whose mass growth during the
pre-main sequence (PMS) phase depends on accretion of gas from the
disc. Therefore, the timescale of disc dissipation sets crucial
constraints for models of stars and planets formation.

Over the past decade, many studies have tried to address these issues
observationally. Based on the assumption that a near infrared (NIR) or 
mid infrared (MIR) spectral excess in young stars are the unambiguous
signature of an inner circumstellar dusty disc, as originally suggested
by Lada \& Wilking (1984), these works have looked at the fraction of
objects in stellar systems of various ages that still show NIR and MIR
excess in their  spectra (e.g. Haisch et al. 2001; Sicilia--Aguilar et
al. 2006; Bouwman et al. 2006; Hernandez et al. 2008). This analysis
suggests a rapid decline of the disc frequency as age proceeds, with
50\,\% of low-mass stars losing their inner dust discs within $\sim
3$\,Myr. However, disc evolution is poorly constrained for ages above
$\sim 5$\,Myr, due to small number statistics both in the number of star
forming regions and in the total number of stars studied. A few
well established cases of long-lived dusty discs have been found so far.
These include the double cluster h and $\chi$ Persei, with up to 8\,\%
of its members still showing IR excess at 8\,$\mu$m at an age of $\sim
13$\,Myr (Currie et al. 2007a), and the Scorpius--Centaurus OB
association, with about one third of the stars showing IR excess at
24\,$\mu$m at ages in the range $\sim 10-17$\,Myr (Chen et al. 2011).

Additional uncertainties on the actual disc fraction are introduced by
the effect of nearby massive stars, which are known to cause the
photoevaporation of the circumstellar discs on a more rapid timescale
(e.g. St\"orzer \& Hollenbach 1999; Guarcello et al. 2010b; De Marchi et
al. 2010a; Mann \& Williams 2010). Furthermore, the observations cited in
the previous paragraph do not probe the gas in the discs (except for the
case of h and $\chi$ Persei; see Currie et al. 2007b), which is expected
to account for $\sim 99\,\%$ of the disc mass. Therefore, most of those
measurements cannot provide strong direct constraints on the late phases
of star formation or on mass accretion. 

Fedele et al. (2010) have compared the fraction of actively accreting
stars with spectral type later than K0 in seven nearby clusters and
associations with that of stars surrounded by dusty discs in the same
systems. They found that at any given age the fraction of stars with
ongoing mass accretion (as determined from the properties of the
H$\alpha$ emission line) is systematically lower than that of stars with
NIR or MIR excess, and concluded that the exponential decay timescale
for mass accretion ($2.3$\,Myr) is even shorter than the timescale for
the dissipation of dusty discs mentioned above (3\,Myr). We note,
however, that this analysis does not explicitly take into account the
effects of temporal variability of the H$\alpha$ emission line (see e.g.
Johns \& Basri 1995; Alencar \& Batalha 2002; Jayawardhana et al. 2006).
Therefore, it only sets a lower limit to the fraction of actively
accreting stars, but the timescale for the dissipation of gaseous discs
is likely longer than $2.3$\,Myr (see also Bell et al. 2013).

Longer lived gaseous discs are known to exist in well studied older
Galactic objects such h and $\chi$ Persei (Currie et al. 2007b), MP
Muscae (Argiroffi et al. 2007), 49 Ceti (Hughes et al. 2008) and
HD\,21997 (Mo\'or et al. 2011) and appear to be needed to explain the
large number of actively accreting PMS stars with ages well above
10\,Myr  recently discovered in the Magellanic Clouds. Studies of the
regions around SN\,1987A and 30\,Dor in the Large Magellanic Cloud have
revealed populations of PMS objects with considerable H$\alpha$ excess
emission (i.e. an H$\alpha$ equivalent width above 20\,\AA), showing
ages of $\sim 15-20$\,Myr and a spatial distribution markedly different
from that of younger ($\la 5$\,Myr) PMS stars (De Marchi et al. 2010; De
Marchi et al. 2011c; Spezzi et al. 2012). Similarly, NGC\,346 and
NC\,602 in the Small Magellanic Cloud contain a conspicuous population
of $\sim 20$\,Myr old PMS objects with mass accretion rates higher than
$10^{-8}$\,\Msolar\,yr$^{-1}$ (De Marchi et al. 2011a, 2011b, 2013). 

The low metallicity of the Magellanic Clouds could be at the origin of
the elevated mass accretion rates and older PMS ages measured in these
works, as suggested by De Marchi et al. (2010, 2011a) and Spezzi et al.
(2012). However, also in the Milky Way there are examples of active PMS
stars with solar metallicity that are older than $\sim 10$\,Myr. They
were detected through their excess H$\alpha$ emission e.g. in  NGC\,3603
(Beccari et al. 2010; Correnti et al. 2011) and in NGC\,6167 (Baume et
al. 2011). These relatively massive clusters host also stars of spectral
type B or earlier, revealing an environment of intense star formation,
more similar to that of the Magellanic Clouds clusters mentioned above
than to the very nearby regions of diffuse star formation studied e.g.
by Haisch et al. (2001) and Fedele et al. (2010). Therefore, it is fair
to wonder whether the paucity of PMS stars with dusty discs or active
mass accretion for ages older than $\sim 5$\,Myr measured by these
authors could be affected by small number statistics, by uncertainties
in the membership selection and perhaps more importantly by different
environmental conditions.

A very interesting object in this respect is NGC\,6611, at the centre of
the Eagle nebula (M\,16) some $1\,750$\,pc away, whose stellar
populations have been recently the subject of a ground- and space-based
multi wavelength study presented in a series of papers by Guarcello et
al. (2007; 2009; 2010b). Using a number of reddening-free colour indices
based on panchromatic observations at visible and infrared wavelengths,
these authors have searched for and identified stars with infrared
excess and have derived their physical parameters. In light of the
appreciable infrared excess, these objects have been classified as
candidate disc-bearing PMS stars. Yet a conspicuous number of them 
appear relatively blue in optical colours and, in the ($V$, $V-I$)
colour--magnitude diagram (CMD), they are consistent with PMS
evolutionary ages in excess of $\sim 10$\,Myr. Recently, medium
resolution spectroscopic observations ($R \simeq 17\,000$) have been
carried out for 20 of these objects by Bonito et al. (2013). Analysis of
the H$\alpha$ line profiles and of their temporal variations show that
10 of these stars (50\,\%) are confirmed members and an additional six
(30\,\%) are likely members, with typical line widths of $\sim
200$\,km\,s$^{-1}$. This shows that accretion/outflow processes are at
work and that these objects must be still relatively young. The question
to answer is how young.

In this work, we  show that their spatial distribution, their
X-ray luminosities, their optical brightness and their positions in the
CMD indicate that most of these stars are at least 8 times older than
the $\sim 1$\,Myr-old PMS population already known in this  cluster. The
structure of the paper is as follows: the data at the basis of the
photometric catalogue are discussed in Section\,2; the stars with
peculiar  blue optical colours are described in Section\,3, where we
also set constraints on the correction for interstellar reddening.
Sections\,4 and 5 present their radial distribution and X-ray
luminosities, whereas Section\,6 addresses the effects that a
circumstellar disc could have on their optical colours. Section\,7 is
dedicated to the discussion of these results, while a summary and
conclusions follow in Section\,8.

\section{Data sample}

The data analysed in this work were extracted from the multiband
photometric catalogue of NGC\,6611 and of the surrounding M\,16 cloud
compiled by Guarcello et al. (2010b). The catalogue contains in excess
of 190\,000 sources, in a region of $33\arcmin \times 34\arcmin$ around
the centre of NGC\,6611. Each object is detected in one or more of the
following bands: optical observations in $B, V$ and $I$ from the ground
(28\,827 sources detected, see Guarcello et al. 2007); NIR observations
in $J, H$ and  $K$ from the 2 Micron All Sky Survey (16\,390 sources
detected, see Guarcello et al. 2007) and from the United Kingdom
Infrared Deep Sky Survey (159\,999 sources detected, see Guarcello et
al. 2010b); observations with the {\it Spitzer Space Telescope} IRAC
camera in four bands centered at $3.6, 4.5, 5.8$ and $8.0\,\mu$m
(41\,985 sources detected, see Guarcello et al. 2009); and X-ray
observations with the {\em Chandra X-ray Observatory} ACIS--I camera
(1\,755 sources detected, see Guarcello et al. 2010b). 

In order to select from the catalogue only candidate cluster members
with circumstellar discs, Guarcello et al. (2010b) have looked for
objects with NIR excess emission, using the standard $[3.6]-[4.5]$ vs.
$[5.8]-[8.0]$ colour--colour diagram defined by the Spitzer/IRAC bands
(e.g. Allen et al. 2004), as well as several reddening-free colour
indices involving a combination of visible and NIR bands (for details on
the indices see Guarcello et al. 2007 and Damiani et al. 2006). Note
that, even though Guarcello et al. (2007; 2009; 2010b) identified stars
with excess in the JHK bands using the reddening-free colour indices,
these objects do show excess even in the more canonical colour--colour
diagrams involving just the NIR bands. As extensively discussed in those
papers (and particularly in Guarcello et al. 2009), the use of
reddening-free $Q$ indices provides a very efficient diagnostic for the
identification of Class II sources, particularly of those with moderate
disc inclination to the line of sight, thanks to the inclusion of the
optical and J bands. Conversely, traditional colour--colour diagrams in
the IRAC bands are better suited for the identification of very young
embedded stellar objects or of Class II stars with highly inclined
discs. 

Guarcello et al. (2010b) identified in this way 834 objects in
NGC\,6611, which are classified as bona-fide cluster members with
circumstellar discs. A total of 624 of them are detected in the $V$ and
$I$ bands, with $V$ magnitude uncertainties smaller than $0.1$\,mag and
$V-I$ colour uncertainties not exceeding $0.15$\,mag. These are the
objects whose properties we study in this work. Their coordinates and
photometry are collected in Table\,\ref{tab1}, which is only available
online and is directly extracted from the catalogue of Guarcello et al.
(2010b).

\begin{table}
\caption{Photometric catalogue, listing the ID number, equatorial 
coordinates (J2000), $V$ and $I$ magnitudes with their uncertainties and
whether the objects was detected in the X rays (Y=yes, N=no, while O
indicates that the object  is outside the field covered by the X-ray
observations). The complete  table is only available online.}
\begin{tabular}{cccccccc}
\hline
ID & RA & DEC & $V$ & $\delta V$ & $I$ & $\delta I$ & X \\
\hline
1 & 274.8771 & -13.7581 & 20.95 & 0.08 & 16.89 & 0.01 & O \\
2 & 274.8531 & -13.6117 & 13.95 & 0.01 & 12.64 & 0.01 & O \\
3 & 274.7775 & -13.7602 & 16.85 & 0.01 & 14.37 & 0.02 & Y \\
4 & 274.7183 & -13.7356 & 21.57 & 0.05 & 17.09 & 0.02 & Y \\
5 & 274.4489 & -13.6591 & 14.68 & 0.01 & 13.01 & 0.01 & N \\
\hline
\end{tabular}
\label{tab1}
\end{table}

\section{Blue stars with excess}

\begin{figure}
\centering
\resizebox{\hsize}{!}{\includegraphics[bb=50 10 468 432]{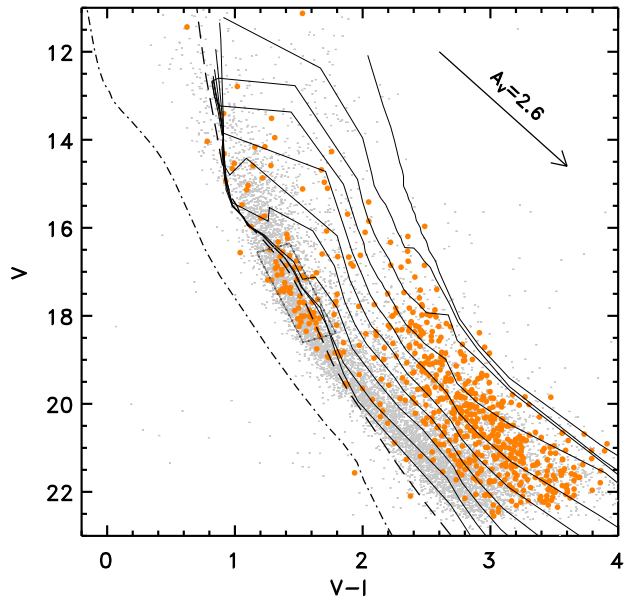}}
\caption{Colour--magnitude diagram of the stars in the field, before
reddening correction. The dots correspond to the 624  disc-bearing stars
with small photometric uncertainties, i.e. less than $0.1$\,mag in $V$
and $0.15$\,mag in $V-I$. Small grey dots show all other stars observed
in this field, with the same selection on photometric uncertainty. The
thin  solid lines are the PMS isochrones of Siess et al. (2000) for
stars of solar metallicity and ages of $0.1, 0.25, 0.5$, 1, 2, 4, 8, 16,
32 and 64\,Myr, from right to left, for a distance modulus
$(m-M)_0=11.2$  and average reddening $A_V=2.6$ (the corresponding
reddening vector is shown by the arrow). The dashed line shows the ZAMS
of Marigo et al. (2008) for stars of solar metallicity and for
$A_V=2.6$, while, for reference, the dot-dashed line shows the same ZAMS
for $A_V=0$.} 
\label{fig1}
\end{figure}

The CMD corresponding to the objects in Table\,\ref{tab1} is shown in
Figure\,\ref{fig1}, where apparent magnitudes before any correction for
extinction are shown. The 624 disc-bearing candidates are indicated with
large dots, whereas with small grey dots we show, for reference, all
other stars detected in $V$ and $I$ in this field, with the same
constraints on photometric uncertainty as for cluster members. Guarcello
et al (2010a) have shown that all these objects are bona-fide cluster
members. However, while most have colours and magnitudes typical of
young ($\sim 1$\,Myr old) PMS stars, about 20\,\% of them appear to have
a $V-I$ colour too blue for their $V$ magnitude and/or to be too faint
in $V$ for their $V-I$ colours for such a young age.

This discrepancy becomes clear when one compares the observed CMD with
theoretical PMS isochrones. The solid lines in Figure\,\ref{fig1}
represent the PMS isochrones of Siess et al. (2000) for stars with solar
metallicity and ages of $0.1, 0.25, 0.5$, 1, 2, 4, 8, 16, 32 and 64\,Myr
(from right to left). All isochrones are reddened adopting the  average
value of the extinction derived by Guarcello et al. (2007) for the stars
in NGC\,6611, namely $A_V=2.6$, and assuming a distance modulus
$(m-M)_0=11.2$. The thick dashed line corresponds to the zero-age
main sequence (ZAMS), from the models of Marigo et al. (2008) for solar
metallicity and the same extinction values, and it agrees well with the
64\,Myr PMS isochrone. The dot-dashed line shows the same ZAMS for
$A_V=0$.  

The peculiarity in this diagram, as noted by Guarcello et al. (2010a),
is the conspicuous group of objects with $16 < V < 19$ and with $1.2 <
V-I < 1.8$. All these stars have NIR excess, as explained above, and as
such are candidate disc bearing objects, but in the CMD they appear
rather blue, i.e. near or at the ZAMS. Hereafter, we will call these
objects ``blue stars with excess'' or BWE for short, following Guarcello
et al.'s nomenclature. For simplicity, all other candidate disc-bearing
objects in the field will be called ``red stars with excess'' or RWE for
short. Although a mismatch between the optical and infrared catalogues
could cause a spurious NIR excess in otherwise normal main-sequence
stars, Guarcello et al. (2010b) have shown that the probability of
mismatches is very low. Furthermore, as mentioned in the Introduction,
Bonito et al. (2013) have shown that the H$\alpha$ emission line profile
of at least half of the BWE stars studied spectroscopically reveals
ongoing accretion. Finally, as we will show in Section\,4, most BWE
stars are located outside of the cluster centre whereas mismatches would
be more likely in the central regions, where the density of optical
sources is higher. 

For the same reason, also a binary origin can be excluded for BWE
stars, since binaries are most likely to be found in the central regions
of clusters (e.g. Sollima et al. 2010). Furthermore, in the photometry
unresolved binaries would appear systematically brighter and redder than
individual components, so we can exclude that BWE objects are binary RWE
stars.

Instead, the positions of the BWE stars in the CMD suggest an older age
than that of the more numerous RWE stars clustered around $V \simeq 20$,
$V-I \simeq 3$ with isochronal ages of order $\sim 1$\,Myr and in any
case less than $\sim 5$\,Myr. The latter correspond to a population of
young PMS stars already known to be present in this field (e.g.
Hillenbrand et al. 1993), while a direct comparison with the PMS
isochrones in Figure\,\ref{fig1} would suggest typical ages in excess of
$\sim 8$\,Myr for BWE objects. In principle, younger ages for the 
BWE stars would still be possible if these objects had a lower
metallicity than that of RWE stars. However, comparison of their colours
with the Pisa PMS evolutionary tracks for a wide range of metallicities
(Degl'Innocenti et al. 2008; Tognelli et al. 2011) reveals that in order
for the BWE stars to have the same age as the RWE stars they would have
to be $\sim 20$ times less metal rich, i.e. they should have $Z=0.001$. 
Star forming regions with such a low metal content are not known in the
Milky Way, nor are such large metallicity differences observed in any
star forming region. Therefore, we will continue to assume that BWE
stars have the same  metallicity as the RWE objects, and as such their
colours indicate an older age.

Even ignoring absolute ages, which may be affected by systematic
uncertainties in the isochrones, Figure\,\ref{fig1} clearly shows that
if the BWE objects are at the distance of NGC\,6611 and are affected by
a similar extinction, they must be considerably older than RWE stars, by
more than an order of magnitude. For instance, using semi-empirical
model isochrones Bell et al. (2013) recently concluded that the most
probable age for young PMS stars in NGC\,6611 is $\sim 2$\,Myr, rather
than 1\,Myr as found by Guarcello et al. (2007), but this would not
change the relative age difference between RWE and BWE stars. In fact,
if PMS ages are systematically underestimated, as Bell et al. (2013)
suggest, the age difference between RWE and BWE stars would grow even
larger.

On the other hand, if patchy interstellar extinction is present inside
NGC\,6611, it is possible that the BWE objects are less reddened than
the other stars and appear older in Figure\,\ref{fig1} just because one
and the same reddening value ($A_V=2.6$) is applied to all isochrones.
Similarly, if the BWE stars are not at the distance of the Eagle Nebula
but somewhere else along the line of sight, they might be bona-fide
young PMS stars. Therefore, it is necessary to understand the presence
and properties of interstellar extinction towards these objects and
where they are located with respect to NGC\,6611, as well as the
effects of field contamination, before we can assign a relative age to
them.

\subsection{Extinction towards the BWE stars}

In their study of NGC\,6611, Hillenbrand et al. (1993) investigated
previous suggestions (Hiltner \& Morgan 1969; Kamp 1974; Neckel \& Chini
1981; Chini \& Kr\"ugel 1983; Chini \& Wargau 1990) that the reddening
law towards NGC\,6611 is different from that of the average
interstellar medium. They concluded that, inside the cluster, the
total-to-selective extinction $R_V = A_V / E(B-V)$ is $3.75$ and as such
it is higher than the canonical interstellar value of $3.1$ (e.g. Mathis
1990). However, they did not find any systematic variation in the value of
$R_V$ or $E(B-V)$ with position in the cluster, contrary to the
suggestions made by Walker (1961) and Sagar \& Joshi (1979) that
reddening in NGC\,6611 increases proceeding westwards and northwards. 

Guarcello et al. (2007) repeated the analysis carried out by Hillenbrand
et al. (1993) using improved photometry and considering separately stars
with and without X-ray emission. They found a somewhat unusual
extinction law, but did not confirm the high $R_V$ value found by
Hillenbrand et al. (1993), concluding instead that $R_V\simeq 3.3$ is
applicable to all objects in the field. In a subsequent paper, Guarcello
et al. (2010b) also derived an extinction map of the region using
bona-fide cluster members. These authors conclude that the reddening is
rather uniform in the central $\sim 10\arcmin$ radius of NGC\,6611,
confirming the average value of $A_V=2.6$ derived before. However, they
also noticed a considerable increase in the reddening north and
northeast of the cluster centre, with local $A_V$ values reaching as
high as 18 mag in regions coinciding with known prominent nebular
structures. 

As we will show in Section\,4, the BWE objects are distributed rather
uniformly over these fields and, as such, only very few of them occupy
the regions with very high extinction to the north and north-east
of the centre. In fact, if they are in the foreground, they might
be affected by lower extinction, as mentioned above. Actually, Guarcello
et al. (2007) already concluded that stars in the foreground of
NGC\,6611 are subject to a lower extinction value, namely $A_V<1.45$,
and the fact that many BWE stars at $V\simeq 17$ appear bluer than the
reddened ZAMS supports this hypothesis.

\begin{figure}
\centering
\resizebox{\hsize}{!}{\includegraphics[bb= 85 190 510 590]{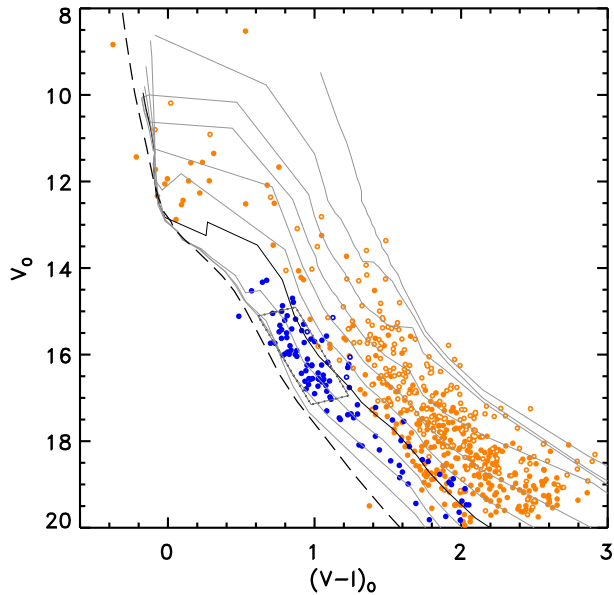}}
\caption{Dereddened CMD of all stars with discs. Light grey dots (orange
in the online version) are RWE objects, to which we have applied a
reddening correction of $A_V=2.6$. Dark dots (blue in the online
version) are BWE stars, to which we have assigned $A_V=1.45$. The solid
lines correspond to the same isochrones as shown in Figure\,\ref{fig1}, for
a distance modulus of $(m-M)_0=11.2$, with the 8\,Myr isochrone shown as
a thicker line. The same distance modulus has been applied to the ZAMS of
Marigo et al. (2008), shown here as a dashed  line. The stars shown as
empty circles are those detected in the X rays. The
meaning of the dotted box around BWE stars is discussed in Section\,7. } 
\label{fig2}
\end{figure}

In Figure\,\ref{fig2} we show a dereddened CMD obtained by applying a
lower reddening correction to BWE objects. For practicality, the acronym
BWE will be reserved for stars with NIR excess (large dots) that in
Figure\,\ref{fig1} are bluer than the 8\,Myr isochrone and with
$V-I<2.6$, whereas all other stars with NIR excess (i.e. candidate
disc-bearing objects) will be labelled RWE. There are in total 513 RWE
and 110 BWE objects.\footnote{In the total count of BWE objects we also
include 29 stars that Guarcello et al. (2010a) had omitted as potential
foreground objects on the basis of their NIR colours.} In
Figure\,\ref{fig2}, we have applied to all RWE objects a reddening
correction corresponding to $A_V=2.6$ and they appear as light grey dots
(orange in the online version), whereas the BWE stars are corrected for
$A_V=1.45$, which as mentioned above is the value measured for
foreground field stars, and appear as dark dots (blue in the online
version). If the BWE stars are associated with the cluster, their
reddening will be larger than  $1.45$, so the adopted $A_V=1.45$ value
should be regarded as a conservative lower limit.

Even with this conservative assumption on the reddening of the BWE
objects, less than {\small $1/4$} of them or 26 out of 110 would appear
younger than the 8\,Myr isochrone (thick line). Most of these 26 stars
would still fall in the age range between 4 and 8\,Myr. This means that
the objects would appear a factor of $\sim 2$ younger than if we had
used $A_V=2.6$ for all stars, which is an uncertainty still consistent
with the formal photometric error of $0.15$\,mag on the $V-I$ colour. In
any case, with isochronal ages in the range 16--32\,Myr, the majority of
BWE stars remain more than an order of magnitude older than the RWE
objects. If $A_V$ were to take on a more realistic value intermediate
between $1.45$ and $2.6$, even older ages would result. Conversely, even
with $A_V=0$ only about half of them would appear younger than 8\,Myr.
For objects on the Galactic plane at a distance of $1.75$\,kpc, assuming
a rough 1\,mag\,kpc$^{-1}$ of visual extinction, we could expect
$A_V=1.75$, which is in line with the range mentioned above. We can,
therefore, conclude that if the BWE stars are at a distance comparable
to that of the Eagle Nebula no reasonable assumptions on the
interstellar extinction can bring BWE and RWE stars to share the same
region of the CMD, so they may well belong to different populations with
different ages. 

\subsection{Distance to the BWE stars}

Alternatively, if BWE objects are PMS stars with an age similar to that
of the RWE objects, given their colours they must necessarily be more
massive and hence located further away than NGC\,6611. More generally,
the fact that all BWE stars have NIR excess places rather stringent
constraints on their distance from us. To better clarify this point, we
list in Table\,\ref{tab2} the distances and reddening values at which
the BWE objects could be located for various choices of their ages. We
have taken as a reference point in the CMD the barycentre of the 56 BWE
stars included in the dotted box in Figure\,1, defined as the smallest
parallelogram that contains the majority (i.e.  $50\,\% + 1$) of these
objects. We have compared the colour and magnitude of its barycentre
with the theoretical isochrones of Siess et al. (2000) for ages of 2, 4,
8, 16 and 32 Myr. Table\,\ref{tab2} lists, for each age, the combination
of distance and reddening giving the best match for the Galactic
extinction law. This comparison shows that:

\begin{enumerate}

\item  For ages of 8\,Myr or less, no acceptable combination of distance
and reddening can be found: objects of this type in the foreground of 
NGC\,6611 should be several magnitudes brighter than the BWE that we
observe, whereas if they were located behind the cluster a match could
be found only for values of the reddening much lower than that towards
the cluster itself. This is hard to imagine due to the considerable
amount of interstellar extinction (e.g. Marshall et al. 2006)  expected
for objects of low Galactic latitude such as NGC\,6611 ($b=0.8$).
Guarcello et al. (2007) already showed that the nebulosity associated 
with the Eagle Nebula blocks or severely reddens the light of background
field stars, which should otherwise appear in large numbers to the left
of the ZAMS in Figure\,\ref{fig1} if they were less extincted than
cluster stars. Furthermore, even allowing for a lower reddening than
that towards NGC\,6611, the paucity of BWE stars in Figure\,\ref{fig1}
at magnitudes fainter than $V=19$ could not be reproduced for a typical
stellar initial mass function (IMF; e.g. De Marchi et al. 2010b). 

\item
At ages of 32\,Myr (and in general older than $\sim 20$\,Myr) PMS stars 
usually no longer show NIR excess. Assuming, nonetheless, that an age of
this type were appropriate for the BWE stars, it would place these
objects in front of NGC\,6611 but not without inconsistencies: (a) it
would typically require a reddening value higher than that for stars in
the immediate foreground of the cluster (i.e. $A_V > 1.45$); (b) for
distances of less than ~ 1\,500 pc the foreground star forming region
hosting the BWE stars would have to be excessively well aligned along
the line of sight to NGC 6611, even though the two regions are not part
of the same complex, and this would look contrived. 

\item
For ages of 16\,Myr, an excellent fit is found for a distance
corresponding to that of NGC\,6611 and the reddening value derived by
Guarcello et al. (2007) for stars in the immediate foreground of the
cluster (i.e. $A_V=1.45$). For distances of less than 1\,650 pc,
however, the isochrones would be systematically too blue to reproduce
the observed colours of stars brighter than $V=15$, thereby implying a
very unusual IMF.

\end{enumerate}

We conclude, therefore, that the only physically plausible explanation
is that the BWE stars are at the distance of the Eagle Nebula, in close
proximity and most likely in the immediate foreground of the NGC\,6611
cluster, and have ages between $\sim 10$ and $\sim 30$\,Myr. As such,
they may thus represent a generation different from that traced by RWE
objects.  

An additional indication of an older age for the BWE stars comes from
their weak {\em Li} absorption lines. The exact amount of {\em Li}
depletion for stars of about 1\,\Msolar\, like the BWE objects depends
on a number of parameters and assumptions entering  the models (e.g.
Jeffries 2006; Tognelli et al. 2012),  such as the stellar mass,
and in some cases large  variations in abundance are seen even for
objects of the same mass (e.g. Bonifacio et al. 2007). For PMS stars,
however, a lower {\em Li} abundance clearly separates more evolved from
very young objects (e.g. Sestito et al. 2008). Of the 20 stars observed
spectroscopically by Bonito et al. (2013), only five have a {\em Li}
equivalent width $EW(Li) > 100$\,m\AA, with a median value of 324\,m\AA.
For five additional objects veiling might be present, making the {\em
Li} absorption line more difficult to detect. Nevertheless, according to
the strong correlation between $EW(Li)$ and age measured by Grankin
(2013) for PMS stars in Taurus--Auriga, a value of $EW(Li) \simeq
100$\,m\AA\, suggests ages of order $\sim 20$\,Myr for objects of about
a solar mass like those in our sample. At least half and possibly
{\small 3/4} of the BWE objects in the spectroscopic sample have weaker
{\em Li} absorption lines, and must therefore be even older.

\begin{table}
\caption{Best fitting values of the reddening towards the BWE stars as a
function of the assumed distance and age. For most distances and ages in the
range explored no best fit is found, whereas for some the reddening should 
take on an unphysical negative value, indicated between square brackets.}
\begin{tabular}{ccccccc}
\hline
$d$ [pc] & $(m-M)_0$ &        &        & $A_V$        &        &          \\
         &           & 2\,Myr    & 4\,Myr    & 8\,Myr & 16\,Myr& 32\,Myr  \\
\hline
1150 & $10.3$ & $\;\,...$ & $\;\,...$ & $\;\,...$ & $\;\,...$ & $\;\,0.5$ \\
1250 & $10.5$ & $\;\,...$ & $\;\,...$ & $\;\,...$ & $\;\,...$ & $\;\,1.3$ \\
1350 & $10.7$ & $\;\,...$ & $\;\,...$ & $\;\,...$ & $[-0.3] $ & $\;\,1.6$ \\
1450 & $10.8$ & $\;\,...$ & $\;\,...$ & $\;\,...$ & $\;\,0.5$ & $\;\,1.7$ \\
1550 & $11.0$ & $\;\,...$ & $\;\,...$ & $\;\,...$ & $\;\,0.9$ & $\;\,1.8$ \\
1650 & $11.1$ & $\;\,...$ & $\;\,...$ & $\;\,...$ & $\;\,1.2$ & $\;\,1.9$ \\
1750 & $11.2$ & $\;\,...$ & $[-1.5] $ & $[-0.1] $ & $\;\,1.5$ & $\;\,2.0$ \\
2000 & $11.5$ & $\;\,...$ & $[-0.6] $ & $\;\,0.9$ & $\;\,...$ & $\;\,2.2$ \\
2500 & $12.0$ & $[-0.6] $ & $\;\,0.7$ & $\;\,1.6$ & $\;\,...$ & $\;\,...$ \\
3000 & $12.4$ & $\;\,0.5$ & $\;\,1.3$ & $\;\,1.7$ & $\;\,...$ & $\;\,...$ \\
4000 & $13.0$ & $\;\,1.2$ & $\;\,1.6$ & $\;\,2.2$ & $\;\,...$ & $\;\,...$ \\
5000 & $13.5$ & $\;\,1.5$ & $\;\,...$ & $\;\,...$ & $\;\,...$ & $\;\,...$ \\
\hline
\end{tabular}
\label{tab2}
\end{table}

\subsection{Field contamination}

Some contamination by field stars is to be expected in the region of the
CMD occupied by the BWE stars. Taking again the dotted parallelogram in
Figure\,\ref{fig1} as representative of their  location in the CMD, we
have compared the number of BWE stars inside the parallelogram with
those expected by the Besan\c{c}on models of Galactic stellar population
synthesis (Robin et al. 2003) in the same region of the CMD. Inside the
parallelogram we find 1\,444 stars, of which 56 are classified as BWE
objects, whereas in the same region of the CMD the Besan\c{c}on models
predict 367 stars due to field contamination. Thus, about one in four of
the observed stars might not belong to NGC\,6611. In principle, it could
therefore seem reasonable to suppose that all 56 BWE objects inside the
dotted parallelogram are simply field stars. However, as we have shown
above, in order for stars in that region of the CMD to have NIR excess
they must have ages of $\sim 16$\,Myr to within a factor of 2 and must
be located at the distance of NGC\,6611 or at most $\sim 100$\, pc in
front of it. Over the entire field of view of these observations, the
Besan\c{c}on models predict no stars within 200\,pc of NGC\,6611 whose
colours would place them inside the dotted region in Figure\,1. It
appears, therefore, extremely unlikely that the BWE stars be field
objects, also in light of their wide H$\alpha$ emission lines, and we
can conclude that even if the BWE objects are not intimately mixed with
the young RWE members of NGC\,6611, they are nevertheless associated
with the M\,16 complex (a precise determination of the actual
location of these objects inside M\,16 will be provided by future
observations with Gaia; Perryman 2005). In the following sections we
investigate the physical properties of these two classes of stars in
more detail, showing that they are indeed remarkably different.

\section{Radial distributions}

\begin{figure}
\centering
\resizebox{7.5cm}{!}{\includegraphics[bb=60 180 500 590]{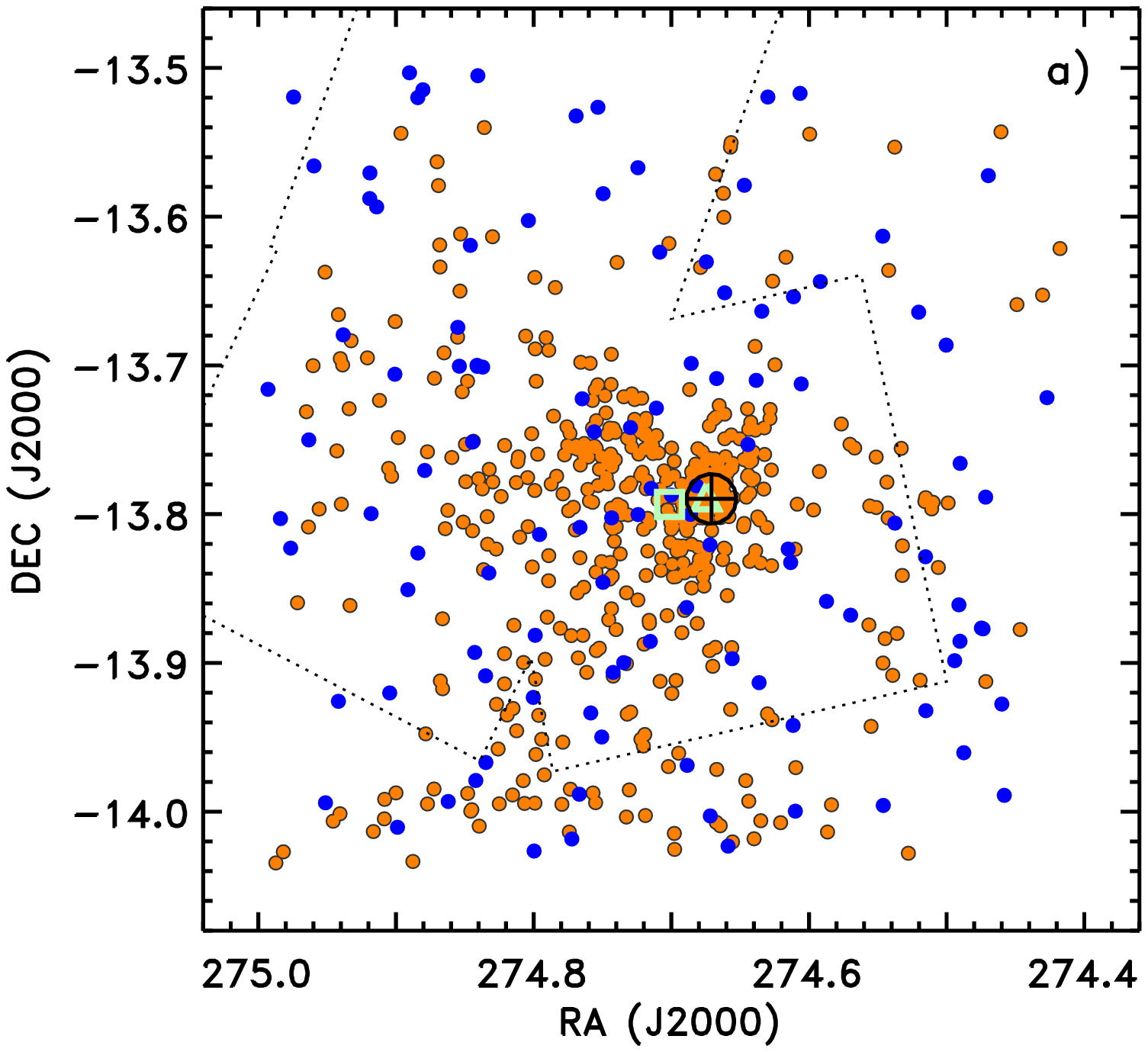}}
\resizebox{7.5cm}{6.5cm}{\includegraphics[bb=60 180 500 590]{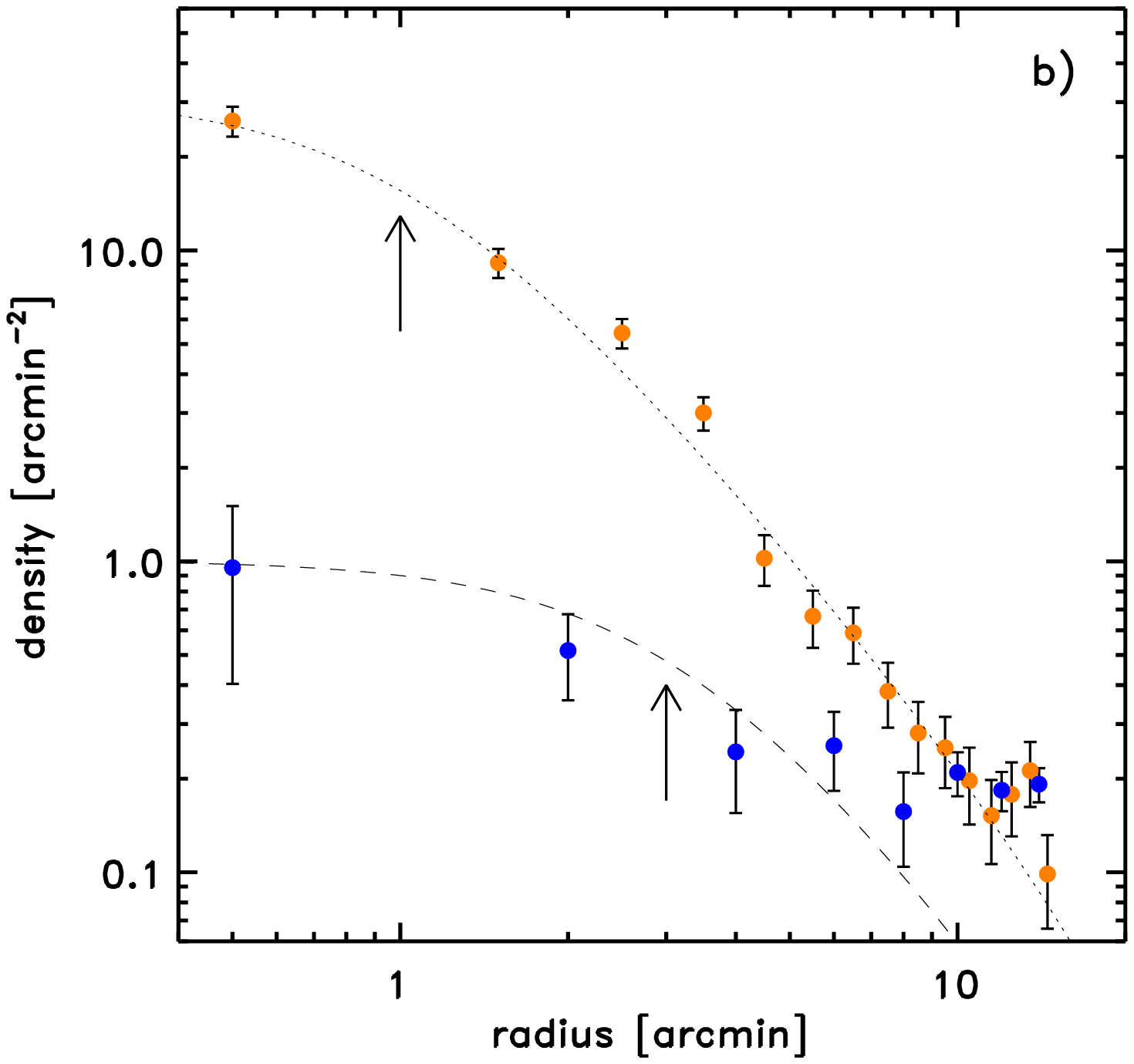}}
\resizebox{7.5cm}{6.5cm}{\includegraphics[bb=60 180 500 590]{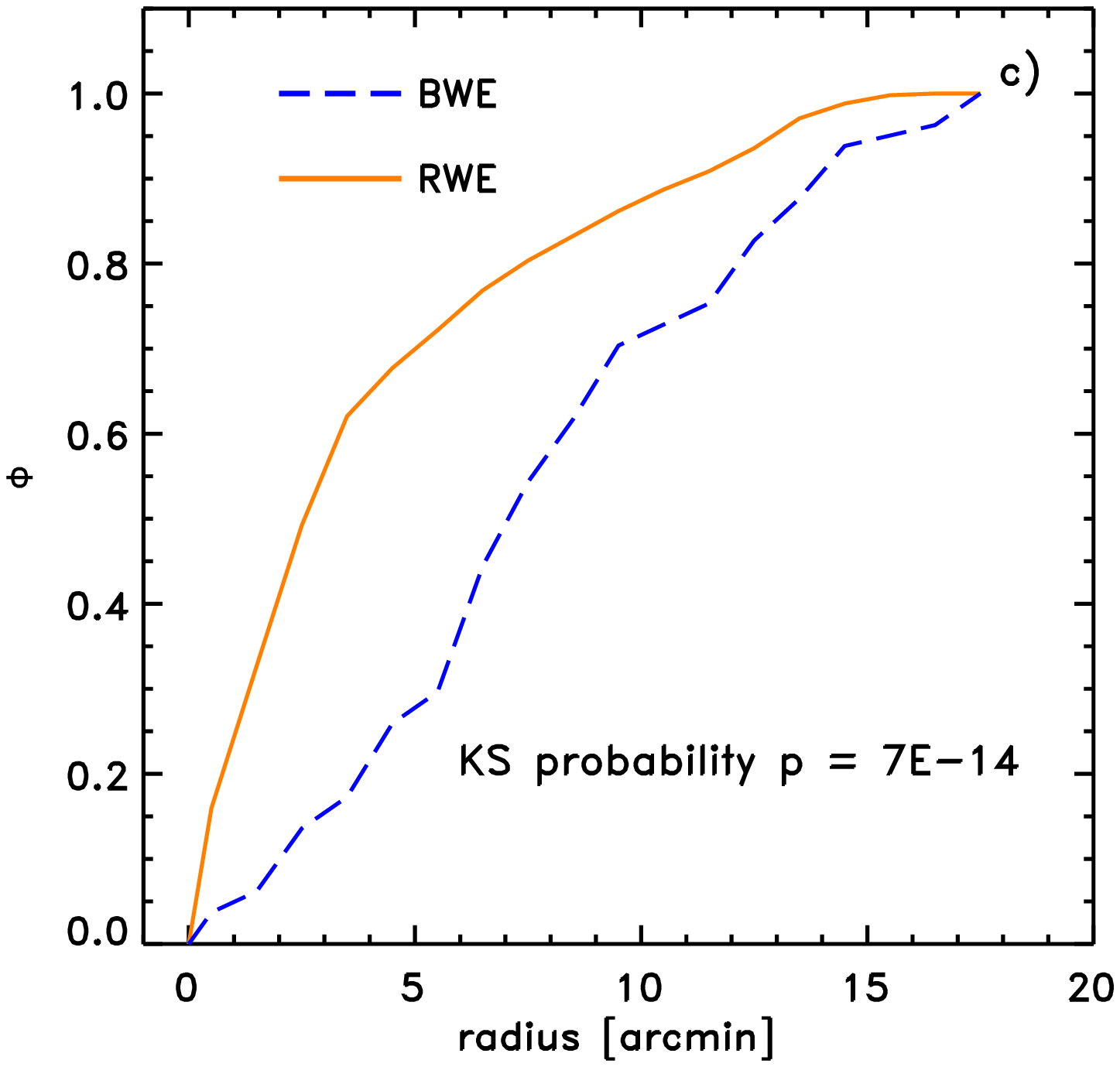}}
\caption{Panel a): locations of BWE and RWE objects, respectively dark
dots (blue in the online version) and light grey dots (orange in the
online version). The nominal cluster centre is the circled cross,  while
the thick square and triangle mark the centre of the distributions of
RWE and BWE stars, respectively. The dotted lines highlight the regions
inside which X-ray observations with Chandra are available. Panel b):
radial density profiles of BWE (dashed line) and RWE (dotted line)
objects. The dashed and dotted lines show the best fitting King
profiles  with core radii of $3\arcmin$ and $1\arcmin$, as indicated by
the arrows. Panel c): cumulative distributions of BWE (dashed line) and
RWE (solid line) objects.}
\label{fig3}
\end{figure}

We first look at the spatial distribution of the two types of objects,
which is shown in Figure\,\ref{fig3}. Guarcello et al. (2010a) had
already noticed that BWE objects are more frequent at larger distances
from the central massive OB stars, but they did not address this issue
in detail. Indeed, the spatial distribution of BWE stars (dark dots in
panel a, shown in blue in the online version) is very different from
that of RWE objects (light grey dots, shown in orange in the online
version). While the latter are strongly concentrated towards the 
nominal centre of NGC\,6611, marked with a circled cross, BWE stars are
much more uniformly distributed across the field.

This is revealed more clearly by Figure\,\ref{fig3}b, which shows the
actual radial density distributions of the two classes of stars. Both
distributions have a higher density near the cluster centre. In
particular, the positions of the two barycentres (shown by the thick
square and triangle in Figure\,\ref{fig3}a, respectively for RWE and BWE
objects) agree to within $\sim 1\farcm5$ of each other and of the
cluster centre. On the other hand, in Figure\,\ref{fig3}a, BWE stars are
clearly less concentrated and more uniformly distributed. 

In order to characterise this difference in a more quantitative way, we
have fitted the distributions with King profiles (King 1966), shown by
the thin dashed and dotted lines. We find core radii $r_c \simeq
1\arcmin$ ($\sim 0.5$\,pc) and $r_c\simeq 3\arcmin$ ($\sim 1.5$\,pc),
respectively for RWE and BWE objects. The $r_c$ value quoted for BWE
stars is actually a lower limit, since the King model underestimates the
profile at large radii.
The two distributions have a common centre (to within $\sim 1\farcm5$),
but  different concentrations and density profiles. A
Kolmogorov--Smirnov test (see Figure\,\ref{fig3}c) shows that the
cumulative radial distributions of the two populations are significantly
different, with a probability of less than one part in $10^{13}$ that
they are drawn from the same distribution.

The fact that BWE and RWE objects not only occupy different regions of
the CMD (see Figure\,\ref{fig1}) but also have a different spatial
distribution suggests that they do not belong to the same population.
Furthermore, from the analysis of their radial distributions, it appears
that the RWE and BWE objects are compatible with having formed virtually
at the same place, and possibly even with a similar initial structure.
This might suggest that the BWE objects have undergone an expansion and
that they are therefore older. In this case, the distribution of 
BWE stars over the area covered by these observations ($33\arcmin \times
33\arcmin$) would be consistent with an expansion velocity of $\sim
1$\,km~s$^{-1}$ (a typical value in Galactic star forming regions) for
a  period of $\sim 15$\,Myr. Although this is not an independent age
determination for the BWE stars, it shows that their spatial
distribution is consistent with an age older than that of RWE stars. An
older age for BWE stars is also supported by the X-ray properties of
these objects, as we show in the following section. 

\section{X-ray luminosities}

An elevated X-ray luminosity ($\log L_X \ga 30$\,erg~s$^{-1}$) is a
characteristic feature of young stars. Stellar X-ray emission from
low-mass objects appears to reach a peak a few Myr after their birth
during the early PMS phase (e.g. Preibisch \& Feigelson 2005; Prisinzano
et al. 2008), then decays slowly as $\sim\,t^{-3/4}$ until $\sim
1$\,Gyr (Preibisch \& Feigelson 2005), and more rapidly thereafter
(e.g. Feigelson et al. 2004). 

Thus, if BWE and RWE objects have systematically different ages, their
X-ray luminosities should also be different. As Table\,\ref{tab1} shows,
there is indeed a remarkable difference in the X-ray properties of the
two types of objects: while only less than 6\,\% of the BWE stars (four
objects in total within the dotted contour in Figure\,\ref{fig3}a) are
detected in the X rays, for the RWE objects this fraction is 46\,\% (or
203 stars). This can be seen in Figure\,\ref{fig2}, where the objects
with an X-ray detection are shown as empty circles. That figure also
shows that the four objects are systematically on the red side of the
BWE distribution, suggesting that they are amongst the younger members
of the BWE population. Figure\,\ref{fig2} would suggest an age of $\sim
8$\,Myr for two of them and less than $\sim 4$\,Myr for the two other
objects, which would make them the youngest members of the BWE group.

Preibisch \& Feigelson (2005) conducted an extensive study of the X-ray
luminosity of PMS stars in the Orion Nebula Cluster (ONC). In spite of
the considerable scatter of $L_X$ (about a factor of 3) at any given
age, these authors found that the median X-ray luminosity of stars in
the mass range $0.5 - 1.2$\,\Msolar, like those considered here, drops
below  $\sim 10^{30}$\,erg\,s$^{-1}$ at ages older than $\sim 10$\,Myr.
The  X-ray detection limit in the study by Guarcello et al. (2010b), on
which Table\,\ref{tab1} is based, corresponds to $\log L_X =
29.8$\,erg\,s$^{-1}$ for the assumed distance of $1.75$\,kpc to
NGC\,6611, so the paucity of X-ray sources in the BWE objects indicates
that most of them are considerably older than $\sim 10$\,Myr. Note
that, in principle, BWE stars could have lower accretion  rates, which
in turn appear to be linked to lower X-ray emission in disc-bearing
stars (e.g. Flaccomio, Micela \& Sciortino 2003). However, this does
not imply that BWE stars are simply RWE stars at the low-end of the
accretion rate distribution, as the rather different spatial distribution 
of BWE stars would ortherwise remain unexplained. 

The distribution of X-ray luminosities in the ONC can be taken as a
reference for NGC\,6611 since the median age and overall X-ray
properties of the two regions are very similar (Guarcello et al.,
2012). Assuming that the relative shape of the distribution does not 
change over time other than for a downward shift in its peak luminosity, 
we conclude that a drop from 46\,\% to 6\,\% in the number of objects
above the detection limit requires a drop a factor of $\sim 4$ in the
median X-ray luminosity. If the two BWE stars with X-ray emission
younger than 4\,Myr are excluded, the fraction is reduced to $\sim
3$\,\% and the corresponding drop in luminosity grows to a factor of
$\sim 6$. These values of the drop can be converted into an age difference
by direct interpolation in Figure 4 of Preibisch \& Feigelson (2005) for
stars of $0.5 - 1.2$\,\Msolar. This gives us an age definitely older
than that of Chamaeleon (7\,Myr) but considerably younger than that of
Pleiades (80\,Myr), of about 12\,Myr in the first case and $\sim
16$\,Myr in the second. These are clearly approximate values, 
but they are in good agreement with the isochronal ages derived from
Figure\,\ref{fig1} for the bulk of the BWE objects.

\section{Effects of disc inclination}

In Section\,3 we showed that, regardless of the extent of the correction
for interstellar extinction, the $V-I$ colours of BWE stars cannot be 
brought into agreement with those of RWE objects. In their study of
NGC\,6611, Guarcello et al. (2010a) looked at other effects that could 
make the young RWE stars appear bluer than their intrinsic photospheric
colours. One of them is intense optical veiling caused by the accretion
process, but the spectro-photometric analysis carried out by these
authors allows them to exclude veiling as the cause of the very blue 
colours for all but a handful of the BWE stars. The recent spectroscopic
study of 20 of these BWE stars conducted by Bonito et al. (2013)
confirms that some veiling is possibly present only in five objects and
is strong for only one of them. A more critical issue considered by
Guarcello et al. (2010a) is that the presence, geometry and inclination
of a circumstellar disc could appreciably alter the apparent optical
colours of the stars, making them considerably bluer than their
intrinsic photospheric colours. Guarcello et al. (2010a) concluded that
no more than {\small 1/3} of the BWE stars could be objects of this
type. In the following we address this effect in more detail and show
that the fraction cannot be larger than a few percent.

It is well known that, in general, reflection nebulae appear bluer than
the illuminating stars because the scattering cross section of grains is
higher at shorter wavelengths (e.g. Henyey \& Greenstein 1938). For the
same reason, if a conspicuous fraction of the light that we receive from
a PMS object is radiation scattered by the circumstellar disc, its
optical colours could be bluer than those of the stellar photosphere and
could make the object appear older in the CMD. This in turn requires the
bulk of the optical photospheric flux to be blocked (i.e. absorbed and
hence reddened) by an almost edge-on circumstellar disc, which makes at
the same time the object considerably fainter. We clarify this scenario
first with a simple order-of-magnitude calculation and then we confirm
its validity using models of the spectral energy distribution of PMS
stars.

\subsection{Order of magnitude estimate}

\begin{figure}
\centering
\resizebox{\hsize}{!}{\includegraphics[bb=110 270 460 735]{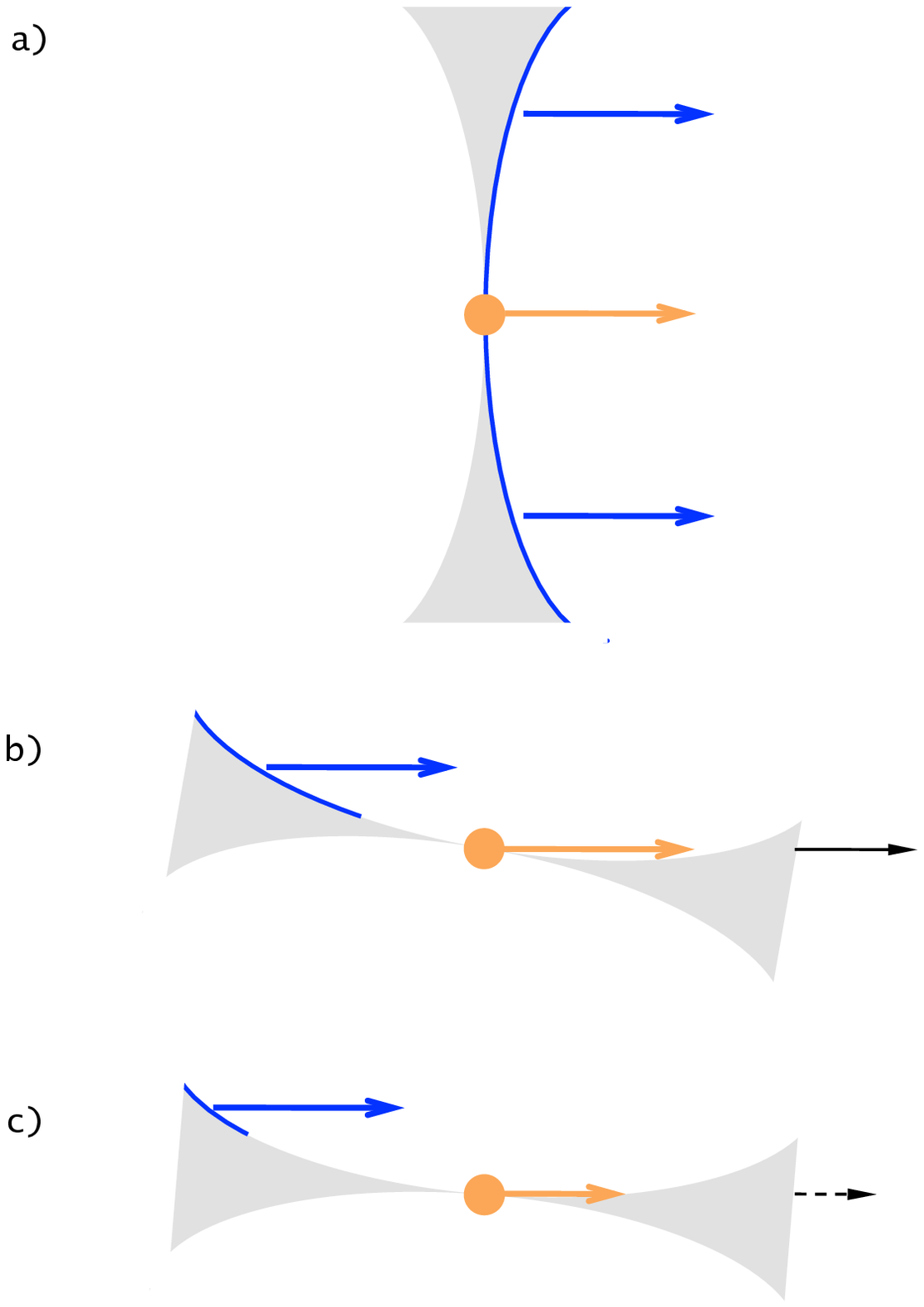}}
\caption{Panel a): if the disc is seen face on, the observer (on the
right) can only see the light scattered by one side of the disc (thick
line, shown in blue in the online version). Panel b): at higher
inclination, a point is reached when the light scattered by the
near side of the disc no longer reaches the observer. Panel c): when 
the inclination increases further, the light of the star will be
absorbed (and reddened) by the near side of the disc and only a portion
of the far side of the disc can effectively scatter light towards the
observer.}  
\label{fig4}
\end{figure}

In the approximation of an infinite disc radius, a flat disc can
intercept {\small $1/4$} of the stellar luminosity $L_\star$ (e.g.
Lynden--Bell \& Pringle 1974). For a flared disc this fraction can be
higher, but also the effect of obscuration will increase with disc
thickness, so the approximation remains valid.\footnote{An opaque disc
is assumed, since for an optically thin disc these effects are much less
important.} Each face of the disc intercepts half of this radiation, or
$L_\star/8$ (see Figure\,\ref{fig4}a), but the fraction that can
effectively be scattered towards an observer seeing the disc face on is
less, since the albedo of astrophysical dust in the optical is typically
less than $\sim 0.5$ (e.g.  Mathis et al. 1983; Natta \& Panagia 1984).
Thus, the fraction of scattered stellar light received by an observer
seeing the face-on disc never exceeds $L_\star /16$, i.e. 3\,mag fainter
than the star itself. Therefore, as long as the disc does not block the
direct view to the star, the preferential scattering of blue light
cannot significantly alter the $V-I$ colour of the object. 

As the inclination of the disc increases, a point is reached when the
light scattered by the near side of the disc no longer reaches the
observer, causing a drop of up to another factor of 2 in the received
scattered flux, or $\sim L_\star/32$ (see Figure\,\ref{fig4}b). When the
inclination increases further, the light of the star will start to be
blocked (i.e. absorbed and therefore also reddened) by the near side of
the disc, but at the same time also a portion of the far side of the
disc is partly blocked, namely the portion closest to the star, thereby
further decreasing the fraction of stellar light effectively scattered
towards the observer (Figure\,\ref{fig4}c). In the case of a perfectly
edge-on disc, that fraction drops to zero for the opaque disc
considered here, but in a more general case we can assume a drop by a
factor of at least two. This brings the total amount of light reflected
by the disc towards the observer to $\la L_\star/64$ and, with the star
no longer directly visible (i.e. completely attenuated by extinction).
It is only at this stage that the colours of the object at optical
wavelengths will appear significantly bluer than those of the
photosphere. However, it must be clear that at this point the brightness
of the object is considerably less than when it is seen at low
inclination, i.e. about $4.5$\,mag fainter. 

In summary, this order of magnitude example shows that while at optical
wavelengths the light received from a star may be made bluer than the
intrinsic photospheric colours, the main effect of increasing the disc
inclination is to make the star appear much fainter and progressively
redder, because of extinction, before becoming blue. Therefore, if the
BWE objects in Figure\,\ref{fig2} were RWE stars with discs seen at high
inclination, they should be considerably fainter than the much more
numerous RWE stars seen at low inclination. As Figure\,\ref{fig2}
already suggests, there are simply not enough bright RWE objects for
this to be the case. This deficiency is quantified in the following
section.

\subsection{Detailed model analysis}

Robitaille et al. (2006) have computed a grid of radiation transfer 
models of young stellar objects (YSOs), covering a wide range of stellar
masses, evolutionary stages, as well as envelope and disc parameters.
The latter take into account theoretical and observational constraints
on the physical conditions of discs and envelopes in YSOs, including
parameters such the  outer radius, accretion rate and opening angle of
the envelope, the mass, inner radius and flaring angle of the disc (see
Robitaille et al. 2006 for details). These models provide the spectral
energy distributions for $\sim 20\,000$ models, each of which is
computed for ten different viewing angles (i.e. disc inclinations), for
a total of over 200\,000 different realisations. These models can be
used to illustrate numerically how the effects of disc inclination
progressively alter the colour and magnitude of stars in the optical
CMD.

\begin{figure*}
\centering
\resizebox{0.95\hsize}{!}{\includegraphics{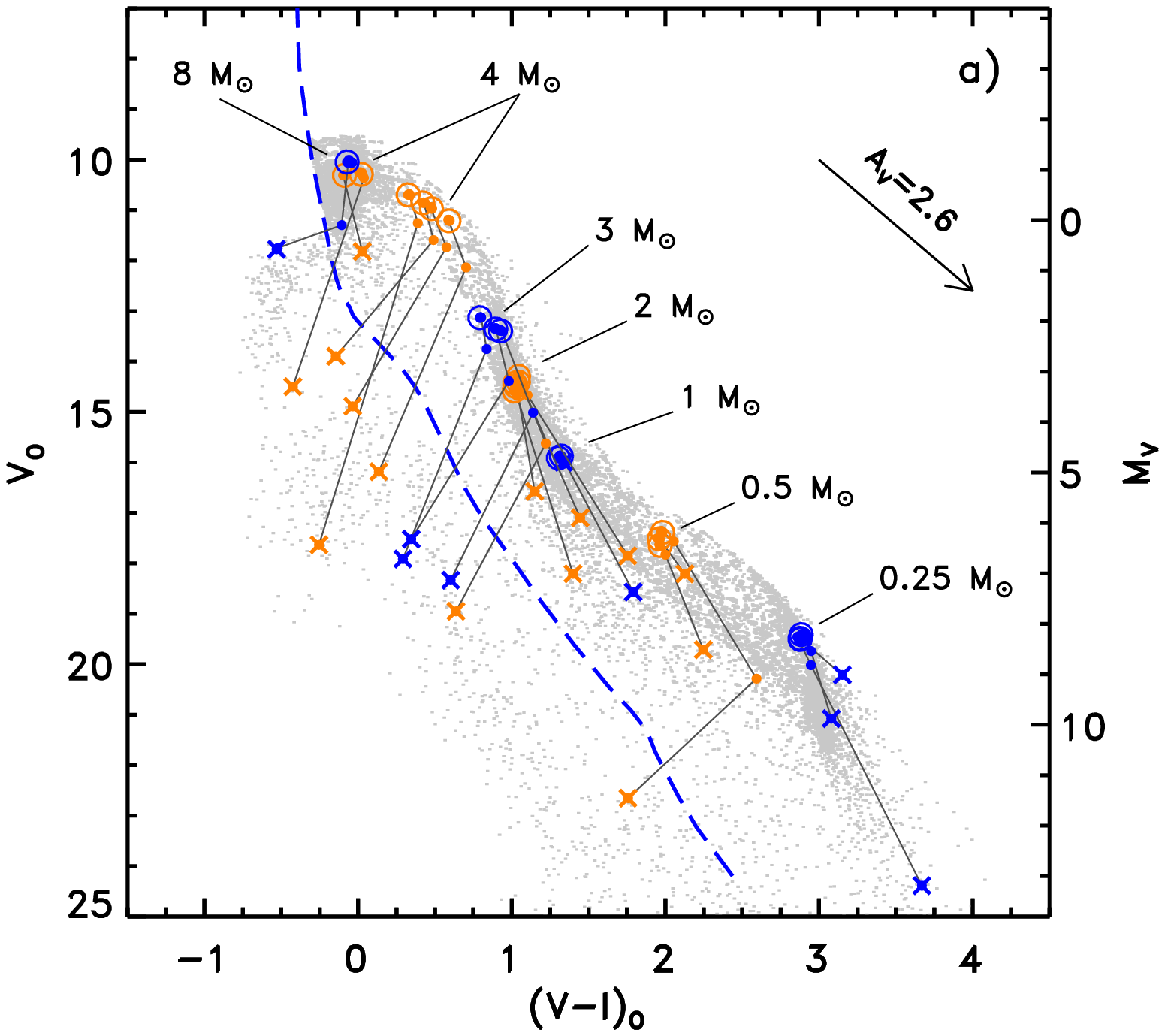}
                      \includegraphics{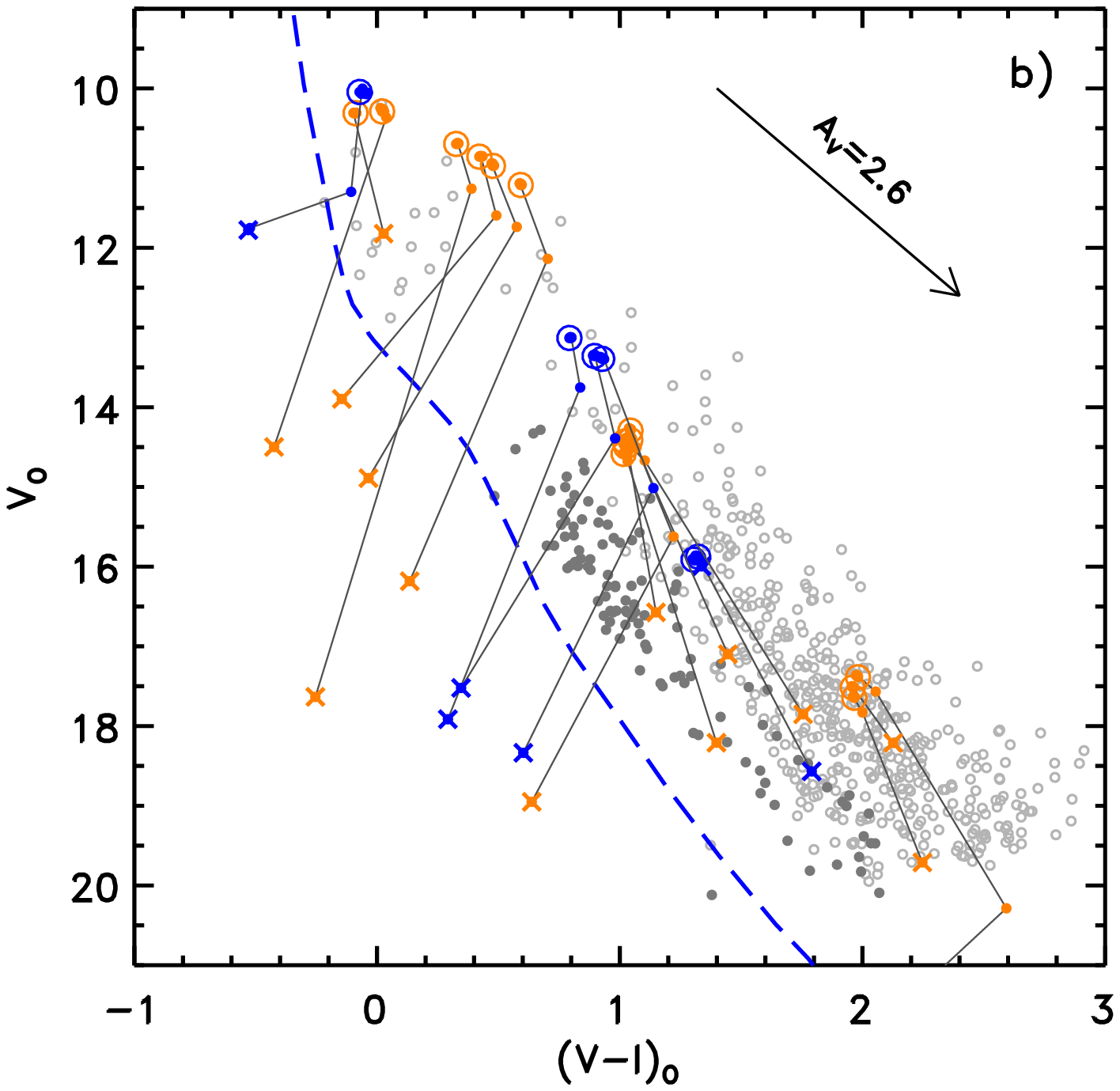}}
\caption{Panel a): Synthetic CMD obtained from the models of  Robitaille
et al. (2006) for an age of $\sim 1$\,Myr and distance modulus $11.2$. The
grey small dots show all $\sim 30\,400$ model realisations with ages in
the range $0.3 - 2.6$\,Myr and masses less that 8\,\Msolar, whereas the
thick symbols correspond to seven representative mass classes extracted
from the grid, as indicated. For each mass class, we show the
displacement of the models as a function  of the inclination of the
discs from almost face-on (large circles, $\theta=18^\circ$) to almost
edge-on (crosses, $\theta=87^\circ$).  
discs are  systematically much  
objects  with face-on discs. Panel b): The models of panel a) are
compared with the CMD position of BWE stars (dots) and RWE stars
(circles). The dashed line in both panels is the ZAMS of Marigo et al.
(2008) as in Figure\,\ref{fig2}.}  
\label{fig5}
\end{figure*}

We have selected from the grid of models a representative sample of
objects with masses of $0.25$, $0.5$, 1, 2, 3, 4 and 8\,M$_\odot$ and an
age of 1\,Myr, since this is the median age of NGC\,6611 as determined
by Guarcello et al (2010b).  Note that since the grid is produced using
a Monte Carlo radiation transfer code, it is usually not possible to
select objects with a specific mass or age, so we have extracted all
objects with ages within 1\,\% of 1\,Myr and with masses within 20\,\% 
of the stated values. These small age and mass dispersion have no 
effects on our conclusions and, instead, they allow us to probe
different regions of the parameter space, e.g. different disc opacities.
As mentioned above, for each model the grid of Robitaille et al. (2006)
provides the spectral energy distribution, including broadband colours,
for ten different viewing angles, from almost pole-on
($\theta=18^\circ$) to almost edge-on ($\theta=87^\circ$), in equal
intervals of cosine of the inclination angle $\theta$. 

In Figure\,\ref{fig5}a we show where these objects are located in the
CMD and how their colours and magnitudes change with disc inclination
because of the effects of disc absorption (i.e. reddening) and
scattering. The ordinates on the left-hand side refer to a distance
modulus $m-M=11.2$, as appropriate for NGC\,6611, while the ordinates on
the right-hand side provide the absolute $V$-band magnitude of the
objects, $M_V$. The small grey dots show all model realisations for
objects less massive than 8\,\Msolar\, and with ages in the interval $0.3
- 2.6$\,Myr, which is the age range derived by Guarcello et al. (2010b)
for this cluster. There are a total of $\sim 30\,400$ models satisfying
these conditions in the grid. The thick symbols in Figure\,\ref{fig5}a
identify the seven representative mass classes  mentioned above. The
large circles correspond to the lowest inclination angle
($\theta=18^\circ$) and the crosses to the highest ($87^\circ$), with
intermediate angles indicated by dots. Note that in most cases
only the point corresponding to the next-to-last angle can be seen, i.e.
$\theta=81^\circ$, since for all others the colour and magnitudes are
practically indistinguishable from those of the face-on configuration.
The dashed line in Figure\,\ref{fig5}a indicates the ZAMS from Marigo et
al. (2008) for solar metallicity, whereas the arrow shows the direction
of the reddening vector for $A_V=2.6$. 

As mentioned in the previous section, when the inclination angle
$\theta$ increases, there is in all cases first and foremost a dimming
and a displacement towards redder colours, along the reddening vector.
Only when the angle exceeds $\sim 85^\circ$ may one find a possible
shift to the blue, while for angles smaller than $\sim 85^\circ$ the
$V-I$ colours are systematically redder than those for a face-on
configuration. Note, however, that not even for the highest inclination
is the colour of the object necessarily bluer than at face-on
conditions, since if the opacity of the disc is not enough to attenuate
the radiation of the central star by a large factor, the scattered light
does not dominate the object's $V-I$ colour. Conversely, if the discs
were systematically thicker and more flared than assumed in the models
of Robitaille et al. (2006), the effects of scattering would make these
objects appear even bluer. However, as mentioned above, the parameter
values sampled by these models directly span those determined from
observations and theories, so such thicker discs would be unusual.

Figure\,\ref{fig5}a also confirms that the order-of-magnitude
calculations in Section\,6.1 are correct, in that young PMS objects that
appear near the MS because of scattering from a high-inclination disc
are typically at least $4-5$\,mag fainter than the same objects seen
at a lower ($\theta < 80^\circ$) inclination angle. 

In summary, we find that: (i) the main effect of a disc at high
inclination is to dim the light from the star and to redden its colours;
(ii) only for less than 10\,\% of the cases (corresponding to extremely
high inclinations, i.e. $\theta > 85^\circ$) is it possible that the
observed colour become bluer than that of the star with a disc seen face
on; and (iii) in such rare cases, the ``bluer'' stars will appear as
sources that are several (say, 4--5) magnitude fainter than their
intrinsic photospheric brightness.

Therefore, if the BWE stars in NGC\,6611 were in reality young PMS stars
like the RWE objects that appear blue only because of a high-inclination
disc, the corresponding objects with discs seen at low inclination must
be considerably brighter and, at the same time, much more numerous. This
is because, as mentioned above, for $\cos \theta > 0.1$, i.e. for
angles smaller than $\theta=84^\circ$, the $V-I$ colours are
systematically redder than those for a face-on configuration. Therefore,
the objects for which disc scattering dominates can be at most 10\,\%
of the total population (recall that when the opacity of the disc is low
the model does not veer to the blue in Figure\,\ref{fig5} even for the
highest value of $\theta$).  

The implications of this conclusion can be seen graphically in
Figure\,\ref{fig5}b, where the same models shown in panel a) are
overplotted on the CMD of NGC\,6611. Grey circles are used for RWE stars
and dark dots for BWE objects. The median colours and magnitudes of BWE
stars are respectively $V-I=1.1$ and $V=16.6$, whereas for RWE stars the
corresponding values are $V-I=1.9$, $V=17.8$ and as such they are
typically fainter than BWE stars. This already indicates that BWE stars
are in general not RWE stars with discs seen almost edge on, since if
that were the case, for every BWE stars there should be more than 10 RWE
objects several magnitudes brighter so that the median $V$ magnitude of
RWE stars should be brighter and not fainter than that of BWE objects.

As an example, there is a large group of BWE stars, comprising
approximately 50 objects, with $(V-I)_0 \simeq 1$ and $15 \la V_0 \la
17$. If these were RWE objects seen at high disc inclination, according
to the models there should be in excess of 500 RWE stars with $V-I > 1$
and $V<13$, whereas only a handful of objects are actually seen in that
region of the CMD (note that further increasing the reddening correction
of BWE stars would not reduce this discrepancy). Therefore, from a
statistical point of view, only few of the BWE objects (particularly
amongst those brighter than $V_0 \simeq 17$), can be stars whose light
is affected by a high inclination disc, while for the majority of BWE
objects the position in the reddening corrected optical CMD reliably
reflects photospheric colours.

\section{Discussion}
 
The analysis described in the previous sections allows us to establish
with a high level of confidence that BWE and RWE objects belong to
populations that are physically different. We summarise here the most
important differences, all of which point to a systematic age difference
between the two classes of objects.

\begin{enumerate}

\item 
By definition, BWE and RWE objects are equally likely candidate
disc-bearing objects associated to the Eagle Nebula, but they occupy
different regions of the optical CMD. No reasonable assumptions on the
amount of extinction or on the extinction law can make the colours and
magnitudes of BWE stars agree with those typical of RWE stars.
Furthermore, although the blue $V-I$ colours of BWE stars could in
principle be displayed by young stellar objects with a circumstellar
disc seen at high inclination, this scenario can only apply to a handful
of objects ($10\,\%$ at most), since it requires both a very high disc
inclination ($\theta > 85^\circ$) and high disc opacity. Also optical
veiling, as shown by Guarcello et al. (2010a), can only be invoked in
just a few cases to explain the blue colours of these stars.
Spectroscopic observations confirm that strong veiling is infrequent
(Bonito et al. 2013). For all other BWE objects, the most natural
explanation of their blue colour is an age older than that of RWE stars.
Isochrone fitting suggests an age  more than an order of magnitude older
than that of the young PMS stars first discovered by Hillenbrand et al.
(1993). Assuming the literature age of $\sim 1$\,Myr for the young PMS
stars (e.g. Hillenbrand et al. 1993; Bonatto et al. 2006; Guarcello et
al. 2007, 2010b), for the BWE objects we obtain an age of $\sim 16$\,Myr
to within  a factor of two. 

\item
The radial distributions of RWE and BWE objects are very different.
Although they share the same centre (to within $1\arcmin$), RWE objects
are strongly concentrated while BWE stars are more widely and uniformly
distributed, with a core radius about 3 times larger ($\sim 3\arcmin$
vs. $\sim 1\arcmin$). Note that this also rules out that BWE objects are
due to mismatches between foreground stars and background NIR sources,
since these would be more likely in the central regions of the cluster,
where the density of optical sources is higher. The presence of
accretion witnessed by the wide H$\alpha$ emission lines in at least
half of the BWE objects studied spectroscopically further excludes the
case of mismatches. A Kolmogorov--Smirnov test shows that the cumulative
radial distributions of the two populations of BWE and RWE stars are
significantly different, at the level of $10^{13}$. An older age for the
BWE stars would explain their wider spatial distribution. 

\item
The X-ray properties of RWE and BWE stars are also very different. Only
6\,\% of the BWE  stars are detected in the X rays above the detection
threshold of  $\log L_X = 29.8$\,erg~s$^{-1}$, while for RWE objects
this fraction is 46\,\%. The X-ray luminosity of PMS stars is known to
decrease with stellar age, so an older age for the BWE stars would
explain their lower median $L_X$. Assuming an X-ray luminosity
distribution similar to that of PMS stars in Orion, the remarkably
different fraction of stars with X-ray detection requires an age
difference larger than 10\,Myr, in agreement with the isochronal ages
mentioned above. 

\end{enumerate}

Each of these three major differences represents a necessary condition 
for an age discrepancy between BWE and RWE objects. Taken individually,
they may not be sufficient conditions, but the fact that they occur
simultaneously leaves little doubt that an age difference is effectively
present. Had BWE stars the same age as RWE objects, in order to appear
bluer they would all need to have a high inclination circumstellar disc.
In addition, one should require a correlation of the spatial
distribution of disc inclinations with the X-ray luminosity of these
objects: all of these conditions appear exceedingly contrived.

One could imagine other effects that would change the age estimate
of BWE stars, but none of them can affect the ages of BWE stars without
affecting in a comparable way those of RWE stars. For instance, the
effects of unaccounted binaries on age estimate of PMS stars have been
discussed in detail by Naylor (2009) and Da Rio et al. (2010), who show
that neglecting unresolved binaries may lead to an underestimation of
the age of all PMS stars of a young cluster by a factor of $1.5-2$.

An older age for the BWE objects is also consistent with the recent
spectroscopic observations at intermediate spectral resolution
($R=17\,000$) carried out by Bonito et al. (2013) for a sample of 20
stars with FLAMES at the Very Large Telescope over the range between
6\,470\,\AA\ and 6\,790\,\AA. These authors find that BWE objects
typically lack a strong {\em Li} absorption line at $6\,708$\,\AA, which
is an indicator of youth in PMS stars (e.g. Barrado y Navascu\'es et al.
2004); only five of these objects have $EW(Li)>0.1$\,\AA\, and only
three of them have $EW(Li)>0.3$\,\AA. At the same time, at least half of
the BWE stars exhibit a strong and asymmetric H$\alpha$ emission line,
with absorption features and typical line widths of $\ga
200$\,km\,s$^{-1}$ revealing active mass accretion.\footnote{Note
that, although in principle also interacting binaries would have
similarly prominent $H\alpha$ emission and NIR excess (e.g. Warner
1995), objects of this type are very rare: they represent only $\sim
3$\,\% of all the stars in the IPHAS catalogue of $H\alpha$
emission-line sources in the northern Galactic plane for which
spectroscopy is available (Witham et al. 2008). As such, their
contribution to our sample is negligible.}

Therefore, the most logical conclusion is that the main difference
between RWE and BWE objects is age, with the important consequence that
star formation must have been active in this field over a period of at
least 20--30\,Myr. Guarcello et al. (2010b) show that the stars born in
the most recent formation episode, i.e. the RWE  objects, have a median
age of $\sim 1$\,Myr, ranging from $0.3$ to $2.6$\,Myr. Isochrone
fitting to the BWE objects in the CMD of Figure\,\ref{fig2} suggests an
age of $\sim 16$\,Myr to within a factor of two for those objects, i.e.
ages ranging from $\sim 8$ to $\sim 32$\,Myr for the majority of them.
Note that this age range does not necessarily imply that star formation
has proceeded continuously over that period of time, but it rather
reflects the limitations in our age resolution, since current
uncertainties on the photometry, on the reddening correction and on the
PMS models do not allow us to resolve ages to better than a factor of
$\sim 2$. Thus, it is possible that more than one burst of star
formation had contributed to the population of BWE stars, but we cannot
distinguish them from one another as clearly as we can separate RWE from
BWE objects.

Concerning the chronology of star formation in M\,16, with an age
approximately between 15 and 30\,Myr, the BWE stars must predate the
formation episode that started some $\sim 3$\,Myr to the SE of
NGC\,6611  and that propagated towards NW. Guarcello et al. (2010b) have
tentatively identified this event in the incidence of a giant molecular
shell (Moriguchi et al. 2002), created some 6\,Myr ago by supernova
explosions. With ages of order $15-30$\,Myr, BWE stars clearly predate
these events and they probably have nothing to do with the birth of the
RWE stars between 3 and $0.3$\,Myr ago.

NGC\,6611 is not the only cluster in which multiple generations of PMS
stars are seen. In the Milky Way, recent observations clearly show that
the star formation in NGC\,3603 has slowly progressed over the past
$\sim 30$\,Myr across the entire cluster (Beccari et al. 2010; Correnti
et al. 2012). In the Magellanic Clouds, a similar star formation pattern
has been seen in 30\,Dor and neighbouring regions (De  Marchi et al.
2011c; De Marchi et al. 2010; Spezzi et al. 2012), in NGC\,346 (De
Marchi et al. 2011a, 2011b) and in NGC\,602 (De Marchi et al. 2013).
But  while in all these regions PMS stars (young and old) are identified
thanks to their H$\alpha$ excess emission, in NGC\,6611 the signature of
the PMS phase is the presence of a circumstellar disc revealed by the
NIR excess.

In fact, in the Milky Way, NGC\,6611 hosts what is presently the largest
population of candidate disc-bearing stars older than $\sim
10$\,Myr and as such it allows us to set much stronger constraints on
the disc fraction at this late stage of PMS evolution than it was
possible so far (see Introduction). The oldest cluster in which a proper
measurement (i.e. excluding non detections) of the disc fraction exists
from NIR and H$\alpha$ excess is NGC\,7160, studied by Sicilia--Aguilar
et al. (2006). The disc fraction that these authors derive is actually
an upper limit in the 2--4\,\% range, but it is based on just three
objects and, as such, it is very uncertain. In the case of NGC\,6611,
the observations discussed in this paper provide a much higher
statistical accuracy. 

As mentioned in Section\,3, the Besan\c{c}on models of Galactic
population synthesis (Robin et al. 2003) predict a total of 367
contaminating field stars inside the dotted box in Figures\,\ref{fig1}
and \ref{fig2} over the whole field of view and out to distances of
50\,kpc for the direction of M\,16. Therefore, it appears that of the
1\,444 stars in that region of the CMD (shown in Figures\,\ref{fig1} as
small grey dots), only 1\,077 are likely cluster members.

This means that in NGC\,6611 the disc fraction for objects older than
$\sim 8$\,Myr and younger than $\sim 32$\,Myr (median age $\sim
16$\,Myr) is at least 7\,\%. In fact, this fraction is still a lower
limit since the probable cluster members may include a substantial
number of stars older than 32\,Myr. In any case, a disc frequency of
7\,\% or higher at an age of $\sim 16$\,Myr implies a characteristic
exponential decay timescale for disc dissipation of $\sim 6$\,Myr,
assuming an initial fraction of 100\,\%. This timescale appears
appreciably longer than the 3\,Myr estimated by Fedele et  al. (2010)
from the analysis of a compilation of nearby clusters with  published
disc frequencies based on NIR excess measurements. However, our result
is fully consistent with the upper limits that Fedele et al. (2010)
derived for older regions and with the observations of h and $\chi$ Per
by Currie et al. (2007a) and those of Sco--Cen by Chen et al. (2011)
mentioned in the Introduction. Furthermore, the semi-empirical
PMS isochrones of Bell et al. (2013) also suggest circumstellar disc
lifetimes about twice as long as currently believed.


The apparent difference between our result and that of Fedele et
al. (2010) may be ascribed to small number statistics (both in the
number of clusters and old stars considered in each cluster) in the
sample compiled by those authors. NGC\,6611, instead, being a more
massive cluster, has a large population of PMS stars with NIR excess
(i.e. candidate disc-bearing objects) also at ages in excess of $\sim
10$\,Myr and this permits more robust statistics. Besides h and $\chi$
Per and Sco--Cen just mentioned, a similar case of relatively massive
cluster is Trumpler\,15, whose stars reveal a disc frequency of $\sim
4$\,\% (Wang et al. 2011) at ages ranging from 4 to 30\,Myr, with a
median of 8\,Myr (Tapia et al. 2003).

We cannot of course rule out that the disc fraction that we obtain in
NGC\,6611 would drop slightly if the contamination by foreground and
background stars were much lower than suggested by the Galactic models
of Robin et al. (2003), who indeed warn that at low Galactic latitudes
model uncertainties are larger. But even in the extreme case of no
foreground or background  contamination, the disc fraction in  NGC\,6611
would remain above the firm lower limit of $3.8$\,\%, implying a
characteristic exponential decay timescale of at least $4.9$\,Myr.

There might be other more physical effects causing the observed
differences, which could be related to the different environmental
conditions prevailing in low-mass nearby star forming regions and in
high-mass star clusters elsewhere in the Milky Way and beyond. For
instance, as mentioned in the Introduction, the proximity of massive
OB-type stars is known to enhance the disruption of circumstellar discs
by photoevaporation, an effect which is directly observed also in
NGC\,6611 (Guarcello et al. 2010b), as well as in Orion (e.g. Mann \&
Williams 2010) and in the LMC (De Marchi et al. 2010). This effect
will reduce the observed disc fraction as time goes by, but the denser
interstellar medium needed to form massive stars in these regions implies 
that the efficiency of photoevaporation process is also reduced. 
Therefore, PMS stars formed more than 8\,Myr earlier that have had
enough time to move away from massive objects could have retained a
larger fraction of their discs.

In any case, these observations underline the importance of
investigating the properties of PMS stars not only in nearby regions of
diffuse star formation such as Taurus--Auriga and $\rho$ Ophiuchi, but
also in more distant and massive starburst clusters. This will be
necessary to secure a more complete and realistic picture of the physics
governing the star formation process, if it is true (e.g. Lada \& Lada
2003) that the majority of stars in the Milky Way, including those in
the solar neighbourhood, have formed in rich clusters containing massive
($>8$\,\Msolar) O-type stars.

\section{Summary and conclusions}

We have studied in detail the population of $110$ stars with prominent
NIR excess and rather blue $V-I$ colours, discovered by Guarcello et al.
(2010a) in NGC\,6611, by comparing them to the $\sim 4$ times more
numerous stars with NIR excess and red $V-I$ colours typical of young
PMS stars. The main results of this work can be summarised as follows.

\begin{enumerate}

\item 

We show that, besides occupying different regions in the CMD, the BWE
stars and RWE stars have rather different physical properties. These
include differences in the spatial distribution (with RWE objects being
much more centrally concentrated than BWE stars) and in the X-ray
luminosity (with only 4\,\% of the BWE stars being detected above $\log
L_X=29.8$\,erg\,s$^{-1}$ versus 46\,\% of the RWE objects). Each of 
these three major differences represents a necessary condition  for BWE
objects to be considerably older than RWE stars. The fact that these
conditions occur simultaneously leaves little doubt that an  age
difference is effectively present and that BWE and RWE stars belong to
two distinct classes of objects. An additional indication of an  older
age for the BWE stars comes from their relatively weak or absent {\em
Li}  absorption lines. Through comparison with PMS isochrones for stars
of solar metallicity, we derive ages in the range $8-32$\,Myr for most
BWE stars, with a median age of 16\,Myr. 

\item

Actually, a very small number of BWE stars could in fact be RWE objects
that appear bluer in $V-I$ because of optical veiling or because of a
circumstellar disc seen at very high inclination. However, the largest
majority of them cannot be objects of this type for the following
reasons: {\em 1)} the main effect of a disc at high inclination is to
dim the light from the star and to redden its colours; {\em 2)} only for
extremely high inclinations ($\theta > 85^\circ$), and hence very
rarely, can the observed colour become bluer than that of the same star
seen with a face-on disc; and {\em 3)} in such rare cases, the bluer
stars will appear at least 4--5\,mag fainter than their intrinsic
photospheric brightness.

\item

These observations provide us with the largest homogeneous sample to
date of Galactic PMS stars considerably older than 8\,Myr that are still
actively accreting from a circumstellar disc as witnessed by the
$\sim 200$\,km\,s$^{-1}$ wide H$\alpha$ emission lines in the spectra of
a sample of these objects. Such a large number of disc-bearing
candidates with a median age of $\sim 16$\,Myr in NGC\,6611 allows us to
set a lower limit of 7\,\% to the disc frequency in this cluster, with
high statistical significance. Assuming an initial fraction of 100\,\%
at birth, these values in turn imply a characteristic exponential
timescale for disc dissipation of $\sim 6$\,Myr.

\end{enumerate}

\section*{Acknowledgments}

We are indebted to Jeffrey Linsky, the referee, for insightful comments
and suggestions that have helped us to improve the presentation of this
work. We are also grateful to Thayne Currie for interesting
conversations about the timescale of disc dissipation. NP acknowledges
partial support by STScI--DDRF grant D0001.82435. MGG's research is
supported by Chandra Grant GO0--11040X. RB acknowledges the support of
the Agenzia Spaziale Italiana under contract ASI--INAF (I/009/10/0).

\end{document}